\documentclass[aps,prb,twocolumn,amssymb,floats,epsfig]{revtex4}
\usepackage{graphicx}
\usepackage{epsfig}
\usepackage{amsfonts}
\usepackage{amsmath}
\usepackage{amssymb}
\usepackage{xcolor}
\usepackage{color}

\newcommand{\bra}{\langle}
\newcommand{\ket}{\rangle}

\begin{document}
\title{Enhancement of superluminal weak values under Lorentz boost}
\author{Abhishek Som${^{1,3}}$ and Sourin Das$^{1,2}$}
\affiliation{\small{$^1$Department of Physical Sciences, IISER Kolkata, Mohanpur, West Bengal 741246, India \\
$^3$Department of Physics, Boston University, MA 02215, USA}}

\begin{abstract}
The local group velocity defined as the weak value of the velocity operator in $1+1$ dimensional Klein-Gordon and Dirac theory is studied. As shown by Berry [J. Phys. A $45, 185308 (2012)$], when the pre- and post-selected states for evaluating the weak value are chosen at random from an ensemble of available states, it gives rise to a universal probability distribution for the local group velocity which can have both subluminal and superluminal components. In this work, we explore the possibility of enhancement of the superluminal fraction of this total probability distribution by applying a Lorentz boost and show that it can indeed be enhanced both in the case of  Klein-Gordon and Dirac theories. \end{abstract}

\maketitle
\section{Introduction}
Since the birth of Einstein's special theory of relativity in 1905
\cite{Einstein} 
and  the velocity addition formula obtained therein, it is evident that superluminal velocity for propagation of an optical signal in vacuum is an impossibility. The idea of signal (or more precisely, information) propagating\cite{Brillouin} superluminally together with special theory of relativity will naturally lead to conclusions which indicate violation of causality.  By 1910, Sommerfeld\cite{Sommerfeld} theoretically established the fact that the velocity of wave-front of a square wave propagating through any medium is always equal to velocity of light in vacuum. But this fact does not rule out possibilities for physical disturbances in medium to propagate at speeds higher than that of speed of light in vacuum without violating causality. In principle, the group velocity of light ($v_g$) can exceed velocity of light in vacuum ($c$). Such situations can arise in case of propagation of light in medium with anomalous dispersion \cite{Born} where the refractive index of the medium could show a decrease in its value as the frequency increases in a particular frequency range. Such medium in modern literature is referred  to as fast light optical medium \cite{Stenner}.

Superluminality or superluminal tunnelling times also appear quite generally in physics of quantum tunnelling where reshaping of the wavefunction is primarily responsible for such effects\cite{Chiao}$^,$\cite {Steinberg}$^,$ \cite{Steinberg1}$^,$\cite{Lee}$^,$\cite{Diener}$^,$\cite{Lee1}. These are referred to as `phenomena  of mode reshaping'. It was shown by Aharonov et.al. \cite{aharonovv} that the tunnelling time can be understood in terms of weak quantum measurements and associated `weak value' \cite{Aharonov} of position of a clock which is weakly coupled to the particle, thereby providing an interpretation for the superluminal tunnelling times in terms of weak values. 

Motivated by such connection of superluminality to weak values and in the backdrop of the predictions of the OPERA experiment, Berry et.al.\cite{BerryNutrino} discussed a scenario which could  be related to the observation of apparent superluminal neutrino. It was shown that such observation might be related to weak values.  Later, in a general study of superluminal speeds for (1+1) dimensional  relativistic random waves, Berry\cite{berry} defines a notion of local group velocity in terms of the weak value for the velocity operator. In this paper, we extend Berry's analysis by studying the influence of Lorentz boost on the probability distribution of the weak value of the group velocity operator. In particular, we explore the possibility of observing an enhanced superluminal fraction of the probability distribution by Lorentz boosting to a desired frame of reference with respect to the lab frame where the wave packet was prepared. We perform this analysis for both one dimensional Klein-Gordon(K-G) and Dirac theories. 

The paper is organised as follows. In section \ref{GroupVel}  we provide a brief introduction to weak values and local group velocity.  In section \ref{KGwaves} we calculate the boost-dependence of the superluminal fraction of the probability distribution due to Lorentz boosting of the K-G waves using averaging techniques given by  Berry and Shukla\cite{berry2}. In section \ref{Dwaves} we study the boost-dependence for the superluminal fraction of the probability distribution for Dirac waves. Finally  in  section \ref{Discussion}, we  present  a  semi-analytic understanding of the numerical results obtained in the article.
\vskip -1cm
 \section{Weak value and local group velocity} \label{GroupVel}
Conventional theory of quantum measurement defines the act of measurement as a process which leads to collapse of a quantum state on to one of the eigenstates of the measurement operator. Averaging over statistically large number of such measurements leads to what we call the expectation value. This type of measurement is referred to as the strong measurement.  Alternatively, thoughts have gone in to defining a notion of weak measurement\cite{Clerk},  where the process of measurement does not necessarily collapse the state but, at the same time gleans very little information regarding the state. These measurements are also referred to as unsharp measurements. By detecting the effect of $\hat{A}$ on a state referred to as the pre-selected state and collapsing the system onto a different state called the post-selected state by performing a strong measurement,  Aharonov et.al. define a new value which was the outcome of this sequence of measurements, named the ``weak value'' which is given below \cite{Aharonov}
\begin{equation}
 A_{weak}=Re\frac{\bra post|\hat{A}|\psi \ket}{\bra post|\psi \ket} \label{weak}. 
\end{equation}
Here, $|\psi \ket$ represents the pre-selected state in which the system is initially prepared and $|post\ket$ represents the post-selected state on which the  system's wavefunction collapses after the weak measurement is performed and $\hat{A}$ is an operator corresponding to an observable whose weak value is determined. Unlike the outcome of standard strong measurement which is bounded by the eigenvalue spectrum of the measurement operator, this weak value is not bounded by the range of eigenvalues of the operator and its real part has the interpretation of conditioned average of generalized observable eigenvalues\cite{Dressel}.  Note that the weak value reduces to the standard expectation value when the pre- and post-selected are chosen to be the same state.  Further, one can generate a probability distribution of the weak values if the pre- and post-selected states are randomly chosen from a random ensemble. It was shown by Berry et.al. \cite{berry2} that if the eigenvalue spectrum of the operator in a given range is dense enough then the distribution function for the weak values generated by choosing random pre- and post-selected states leads to universal generalized Lorentzian function, independent of the statistics of the pre- and post-selected states and the exact distribution of the eigenvalues. In this article we will be using such an averaging scheme for weak values. 

Next we consider the notion of local group velocity which can be defined via the operator given by $\hat{v}(x)={1/2} {(\delta(x-\hat{x}) \hat{v} + \hat{v} \delta(x-\hat{x}))}$. It has been shown\cite{berry} that weak values of the group velocity operator, $\hat{v}=\partial_{\hat{k}}{\omega(\hat{k})}$ (here, k is the momentum operator),  evaluated for a pre-selected state taken as a wavepacket composed of several subluminal plane waves and a post-selected state taken as the position eigenstate equals the expectation value of local group velocity operator $\hat{v}(x)$. A distribution function for the weak values can be obtained by choosing random pre- and post-selected states. In what follows we show that for relativistic random waves in 1+1 dimensions,  if the frame of the observer is altered with respect to the frame in which the wavepacket was prepared (we call this the lab frame), then the probability for weak value of group velocity operator going superluminal is also altered significantly either in favour of or against it. The superluminal probability can be defined as: 
\begin{equation}
P_{super}=\int_{-\infty}^{-c} P(v)dv + \int_{c}^{\infty} P(v)dv , \label{total}
\end{equation}
where the probability density of local group velocity is given by $P(v)$.
For the remainder of the paper, we use $\hbar=1$ , $c=1$ and particle mass $m=1$, so that the superluminality condition is $|v|>1$.

\section{ Klein Gordon waves} \label{KGwaves}
We start with the study of superluminality in 1+1 dimensional K-G equation as given below :

\begin{equation}
 \frac{\partial^2 \psi}{\partial t^2} - \frac{\partial^2 \psi}{\partial x^2} = -\psi. 
\end{equation} 

An initial state can be prepared by superposition of plane waves solutions of the K-G equation given by $\psi(x,t)=\sum_{k}c_k \exp{(i\gamma_k(x,t))}$
,where the phases are given by $ \gamma_k=\mu_k +kx-t\sqrt{k^2+1} $. This state will be eventually taken as the pre-selected state for the weak measurement. To define the wave packet $\psi(x,t)$, we need to specify the set of contributing wavevectors k, their real excitation amplitudes $c_k$ and the phases $\mu_k$ ; in later sections, we will choose a certain power spectrum for ${c_k}^2$ , and $\mu_k$ is taken to be a random variable. The angular frequency can be written as: $\omega=\sqrt{k^2+1}$. The standard definition of group velocity $\hat{v}=\partial_{\hat{k}}{\omega(\hat{k})}$ gives
\begin{equation}
  \hat{v}= \frac{\hat{k}}{\sqrt{\hat{k}^2+1}} \label{u1}. 
\end{equation}

Now by choosing the pre-selected state to be the wave packet $\psi(x,t)$ given above and the post-selected state to be the position eigenstates, following Eq.~[\ref{weak}] the weak value for  $\hat{v}$ can be evaluated to be

\begin{equation}
 v(x,t)=Re \frac{\sum_{k} c_k \frac{k}{\sqrt{k^2+1}} \exp(i\gamma_k(x,t))}{\sum_{k} c_k \exp(i\gamma_k(x,t))}.
\end{equation}

We now define a new variable $u\equiv \frac{k}{\sqrt{k^2+1}}$ and  we relabel the following variables as $ c_k \equiv c_u$, $\gamma_k(x,t) \equiv \gamma_u(x,t)$ for convenience of notation in our further calculations. Hence the expression for weak value of  $\hat{v}$ or equivalently the local group velocity can be re-expressed as

\begin{equation}
 v(x,t)=Re \frac{\sum_{u} c_{u} u~ \exp(i\gamma_u(x,t))}{\sum_{u} c_u~ \exp(i\gamma_u(x,t))} \label{localgroup} .
\end{equation}
The probability density of the local group velocity $P(v)$ needs to be determined in order to obtain the superluminal probability given by Eq.~[\ref{total}]:

\begin{equation}
     P(v)=\bra\bra\delta(v-v(x,t))\ket{\ket_{\mu_k}} . 
 \label{delta}
 \end{equation}
Here the notation $\bra \bra..\ket\ket$ stands for statistical ensemble averaging over random values of ${\mu_k}$ distributed in the interval $[0, 2\pi]$. It is important to note that ergodicity is assumed which implies that the ensemble average of the velocity operator over random values of $\mu_{k}$ for a fixed $x, t$ will be the same as averaging over all possible values of $x$ or $t$ while keeping $\mu_k$ a constant. The validity of such ergodicity was checked numerically\cite{berry,berry2}. 

As mentioned earlier, the primary aim of this article is study the influence of Lorentz boost on $P_{super}$ which we are going to explore numerically. We call the frame in which the wavepacket is prepared as the ``lab frame" or ``rest frane'' while the frame of reference after boosting will be called the ``boosted frame".  We follow a convention where the lab frame variables carry the subscript ``lab" while the boosted frame variables carry no subscript. Under the action of the Lorentz boost on the two component wave-vector, it transforms as: $k \rightarrow \gamma (k-  \beta  \omega)$ and  $\omega \rightarrow \gamma (\omega- \beta  k)$, where $v$ is the boost velocity, $k$ is the momentum and $\omega$ is the frequency  and $\gamma=1/\sqrt{1-v^2}$. Hence the corresponding group velocity modifies as  

\begin{equation}
 u_{lab} \rightarrow u = \frac{k}{\omega}= \frac{k_{lab}-\omega_{lab} v}{\omega_{lab} - k_{lab} v} = \frac{u_{lab}-v}{1-u_{lab} v}.
 \label{mapping}
\end{equation}
In the lab frame, the group velocity of the wavepacket post-measurement is (following Eq.~[\ref{localgroup}]):
\begin{equation}
v(x,t)=\Re\left[ \frac{\sum_{u_{lab}} c_{u_{lab}} u_{lab} \exp(i\gamma_{u_{lab}}(x,t))}{\sum_{u_{lab}} c_{u_{lab}} \exp(i\gamma_{u_{lab}}(x,t))} \right]~.
\label{labspeed}
\end{equation}
Using normalization of probability we have  $\int  c_{u_{lab}} ~d{u_{lab}} = \int  c_{u} ~d{u}=1$. Hence the group velocity spectrum in the boosted frame can be expressed as
\begin{equation*}
  c_u = c_{u_{lab}}(u) \dfrac{d}{du} u_{lab}(u)
= c_{u_{lab}}\Big(\frac{u+v}{1+uv}\Big)\dfrac{d}{du}\Big(\frac{u+v}{1+uv}\Big)
\end{equation*}
\begin{equation}
= c_{u_{lab}}\Big(\frac{u+v}{1+uv}\Big)\frac{1-v^2}{(1+uv)^2}
\label{transformation}
\end{equation}

Also, note that the phase in Eq.~[\ref{labspeed}], being a Lorentz scalar, remains invariant under such a transformation. Thus, in the boosted frame, the group velocity in Eq.~[\ref{labspeed}] is transformed to:

\begin{equation}
 v(x,t)=Re \frac{\sum_{u} c_{u} u~ \exp(i\gamma_u(x,t))}{\sum_{u} c_u~ \exp(i\gamma_u(x,t))}
\end{equation}

where $c_u,u$ is given in terms of the rest frame coefficients $c_{u_{lab}},u_{lab}$ as in Eq.~[\ref{transformation}] and Eq.~[\ref{mapping}] respectively.

We analyse a simple wavepacket consisting of a superposition of $N>>1$ identical plane waves, with a double Gaussian spectra centered at $ \mu_{1/2}$ and a standard deviation of $\sigma = \sigma_{1/2}$ given by 
 \begin{equation}
  c_{u}^2= \frac{1}{\sqrt{2\pi\sigma_1^2}}e^{-\frac{(u-\mu_{1})^2}{2\sigma^2}}+\frac{1}{\sqrt{2\pi\sigma_2^2}}e^{-\frac{(u-\mu_2)^2}{2\sigma_2^2}}, 
\label{dist}
 \end{equation}
 
For our numerical analysis we work with a simple wavepacket consisting of a superposition of $N>>1$ identical plane waves, with  $\mu_{1/2}=\pm 0.5c$ and a standard deviation of $\sigma = 0.1$. This wave packet defined in the rest frame will be taken as a reference wavepacket for studying the $P_{super}$ in the rest and the boosted frame. For $N>>1$, the distribution of the group velocities is not relevant\cite{berry2}; whether it is randomly selected or regularly arranged is inconsequential as far as the probability distribution post measurement is concerned. We will elaborate on this later in section\ref{Discussion} in the paper. Now, to obtain the probability distribution for superluminal velocity in the boosted frame numerically we follow the following steps:  
(a) Creating the initial group velocity distribution: For every simulation below, we consider a wavepacket selecting 200 plane waves, 100 of which move right and the remaining 100 left with amplitude given according to the spectrum in double Gaussian spectra mentioned above. We, therefore, simulate a bunch of plane waves with close group velocities moving in opposite directions, the superposition of which is our pre-selected state. Relative phases between these components are chosen from a uniform random distribution on $[0, 2\pi]$.
(b) Calculating the final group velocity distribution: In order to get the post-averaged distribution of group velocities, we consider a certain point in the $x-t$ plane and compute the group velocity at that point when these 200 waves are evolved under the Klein-Gordon equation. This calculation is repeated 5000 times with a different set of randomly chosen 200 phases and a group velocity distribution is obtained. The superluminal fraction is given by the fraction of times the group velocity exceeded unity. For any given value of boost, above described steps are repeated in conjunction with Eq.~[\ref{transformation}]. The superluminal fractions hence obtained are then plotted against the corresponding values of boost. 

Ergodicity ensured that performing an averaging over random phases and  random space time points provides the same result.  Hence  we could have averaged over space-time instead of averaging over phases. We also performed numerical calculations,  where instead of choosing 5000 different sets of 200 random phases, we choose 5000 sets of position or time keeping the set of random phases fixed. Position and time are distributed appropriately such that $kx$ and $\omega t$ are uniformly distributed between 0 and $2\pi$ and we obtained the same set of results within bounds of permissible error. \\

\begin{figure}
 \centerline{\includegraphics[width=10cm]{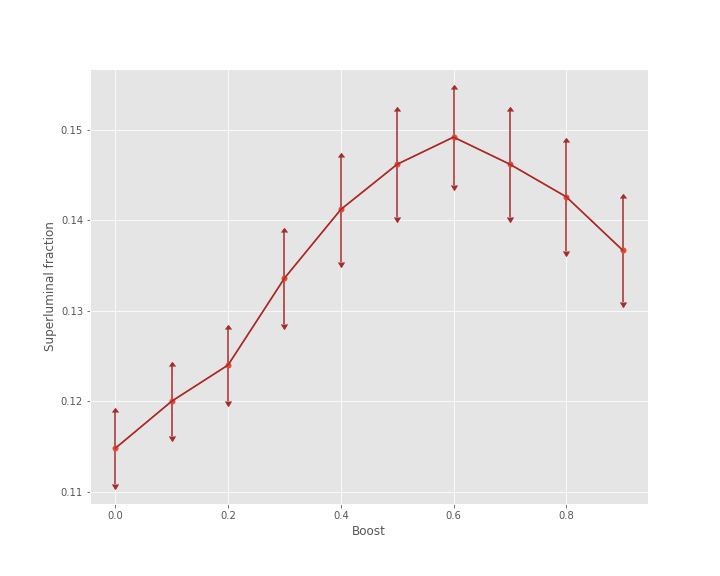}}
 \caption{$P_{super}$  is plotted as a function of the boost velocity ($v$). The distribution in the lab frame comprises of two sharply peaked($\sigma=0.1$) Gaussians with $\mu_{1}=0.5c$ and $\mu_{2}=-0.5c$. Error bars show an average sampling error across all boosts to be of around 4.1\%.}
 \label{KGboost}
 \end{figure}

With the above described scheme for numerical evaluation we explore the possibility of  enhancement of the superluminal probability, $P_{super}$, under the action of boost. We will see that the answer is indeed affirmative and this is the central finding of this paper. We also note that the superluminal probability does increase with boost but eventually it falls off to zero in the ultra relativistic limit i.e., $v\rightarrow c$ (see Fig.~[\ref{KGboost}]). This is interesting as it implies that given a group velocity distribution in the lab frame, there is a value of boost for which the corresponding $P_{super}$ in the boosted frame has large enhancement of superluminality. In the specific case in figure [\ref{KGboost}], the maxima is reached at $v=0.6c$ at which the superluminal fraction is around $32\%$ higher than the lab frame. 
 
\begin{figure}
 \centerline{\includegraphics[width=10cm]{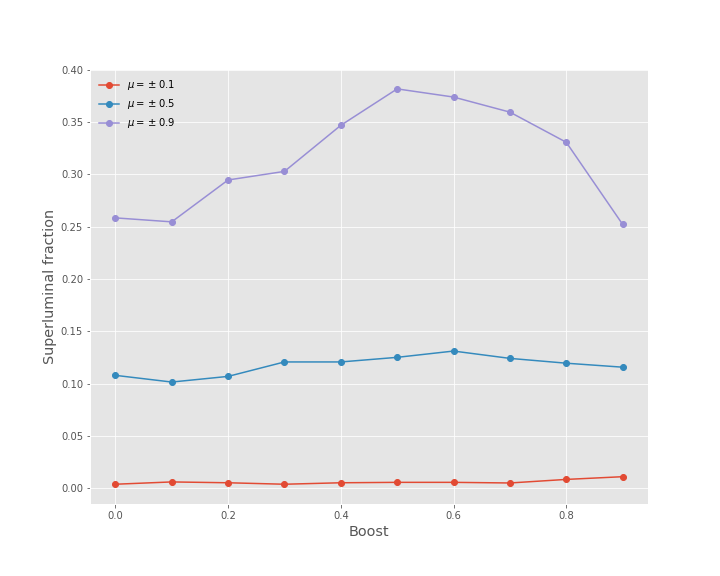}}
 \caption{$P_{super}$  is plotted as a function of the boost velocity ($v$). The distribution in the lab frame comprises of two sharply peaked($\sigma=0.01$) Gaussians with $(\mu_{1},\mu_{2})=\pm 0.1c,\pm 0.5c,\pm 0.9c$ respectively, from bottom to top.}
 \label{symm_compare}
 \end{figure}

\begin{figure}
 \centerline{\includegraphics[width=10cm]{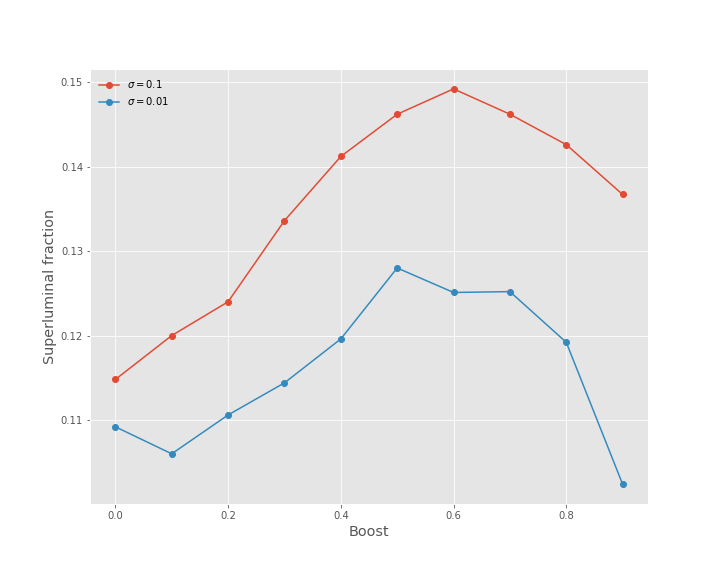}}
 \caption{$P_{super}$  is plotted as a function of the boost velocity ($v$). The distribution in the lab frame for the blue curve comprises of two sharply peaked($\sigma=0.01$) Gaussians with $(\mu_{1},\mu_{2})=\pm 0.5c$ and that for the red curve is made of $(\mu_{1},\mu_{2})=\pm 0.5c$, with a wider Gaussian ($\sigma=0.1$)}
 \label{KGcomp}
 \end{figure}

We study the influence of various parameters in the initial spectrum which could influence the superluminality, i.e., $P_{super}$  in the boosted frame. We show that if we consider a pre-selected 
wavepacket which comprises of more number partial waves which lay closer to the velocity of light in the lab frame, then we end up having a more prominent enhancement of  superluminality in a suitably chosen boosted frame. We compare the variation of $P_{super}$ with boost for three different cases, $\mu = \pm0.1c,\mu = \pm0.5c$ and $\mu = \pm0.9c$ as presented in Fig.~\ref{symm_compare}. We observe that the superluminal fraction in the boosted frame is much higher for spectra corresponding to $\mu =\pm0.9c$ as opposed to $\mu = \pm0.1c$. Next, we numerically study the influence of variance of the spectrum in lab frame on $P_{super}$  evaluated in the boosted frame. Instead of taking a spectrum that is as wide as $\sigma=0.1$, we choose a much narrower spectrum, $\sigma=0.01$,i.e, the waves move with extremely close wavenumbers. The result of this study is presented in Fig.~\ref{KGcomp}. We observed that the reduction of $\sigma$ tend to cause suppression of superluminality. A discussion on the understanding of behaviour of $P_{super}$ as a function of boost is presented later in section~\ref{Discussion}. In the next section, we study $P_{super}$ in the case of Dirac waves as a function of boost. 

\section{Dirac waves}\label{Dwaves}
Dirac equation in 1+1 dimension can be written as : 
\begin{equation}
 i\partial_t |\psi\ket=\hat H|\psi\ket = ({\sigma_z}\hat{k}+{\sigma_x})|\psi\ket , 
 \label{Diraceq}
\end{equation}
where $|\psi\ket$ is a two component column vector and $\sigma_x,\sigma_z$ are  the $x$- and  $z$-components of the Pauli matrices respectively. One can easily relate this to the usual 3+1 dimensional Dirac equation $i\partial_t |\psi\ket= ({\vec{\alpha}}\cdot\hat{p}+\hat{\beta}m)|\psi\ket$ by substituting $\hbar=1$ , $m=1$ and replacing the  Pauli matrices by the corresponding Gamma matrices. Now, rewriting $\psi=(\eta \ ,\ \chi)$,  we get 

\begin{equation}
 \left(\begin{array}{c} i\partial_{t}\eta \\ i\partial_{t}\chi \end{array}\right)
 =\left(\begin{array}{cc} 1 & 0\\ 0 & -1 \end{array}\right)
 \left(\begin{array}{c} -i\partial_{x}\eta \\ -i\partial_{x}\chi \end{array}\right)
 +\left(\begin{array}{cc} 0 & 1\\1 & 0 \end{array}\right)
 \left(\begin{array}{c} \eta \\ \chi \end{array}\right)
 \label{Diraceq1}
\end{equation}
The effect of Lorentz boost is incorporated as $(\eta \ ,\ \chi)\rightarrow (\eta \ ,\ \chi)\hat{U}^T $ where $\hat{U}$=diag$(\sqrt{\frac{1+v}{1-v}}, 
\sqrt{\frac{1-v}{1+v}})$  represents the action of boost on the two component wavefunction  and $v$ is the boost velocity. Thus the boosted spinor is given by $(\eta \ ,\ \chi) = (\sqrt\frac{1+v}{1-v}\eta \ ,\ \sqrt\frac{1-v}{1+v}\chi)$. Next, we note that the group velocity operator for the Dirac equation is given by $\hat v=\partial_{\hat k}\hat H ={\hat{\sigma}_z}$ whose eigenvalue spectrum is bounded by $\pm 1$ in units of velocity of light as expected. In this case also, we will be evaluating the weak values for this group velocity operator by considering  the pre-selected state to be a wavepacket formed by superposition of plane wave solutions of the Dirac equation with random relative phase differences between them while the post-selected state will be taken to be the position eigenstate taken at random positions.  \\
\\
To implement the random choice for the post-selected state, we first note that the normalized two component wavefunctions of the Dirac equation can be mapped on to the Bloch sphere owing to the two component nature of the wavefunction. We choose a uniform random distribution for the  two component wavefunctions  by choosing a uniform random distribution of these states on the  Bloch sphere. The post-selected state is represented by $\bra post| = \left(\begin{array}{cc} \cos(\theta/2) &, e^{-i\phi}\sin(\theta/2) \end{array}\right)$ where $\theta$ and $\phi$ are taken to be uniform random variables. Hence the weak value of the group velocity operator in the boosted frame of reference is given by
 \begin{align*}
  V(x,t) = Re \Bigg[\frac{\cos(\theta/2)\sqrt{\frac{1+v}{1-v}}\eta -  {e^{-i\phi}}\sin(\theta/2)\sqrt{\frac{1-v}{1+v}}\chi}{\cos(\theta/2)\sqrt{\frac{1+v}{1-v}}\eta + {e^{-i\phi}}\sin(\theta/2)\sqrt{\frac{1-v}{1+v}}\chi}\Bigg]
   \end{align*}
  \begin{align}
  = \frac{1-\tan^2(\theta/2)|{\frac{\chi}{\eta}}|^2(\frac{1-v}{1+v})^2}{1+\tan^2(\theta/2)|{\frac{\chi}{\eta}}|^2(\frac{1-v}{1+v})^2 + 2\tan(\theta/2)|{\frac{\chi}{\eta}}|\frac{1-v}{1+v}\cos\varphi} \label{hj} ,
 \end{align}

where $V(x,t)=Re[\bra post|\hat{V}|\psi\ket / \bra post|\psi\ket]$ and $\varphi = \phi - arg(\frac{\chi}{\eta})$. Next, we note the two component wavefunction corresponding to the positive energy eigenstate of the Dirac equation [\ref{Diraceq}] in the lab frame can be expressed  in terms of the group velocity for the partial plane waves ($u=k/\sqrt{1+k^2}$) defined in Eq.~[\ref{u1}] as  $(\sqrt{1+u},\sqrt{1-u})$.  \\

\begin{figure}
 \centerline{\includegraphics[width=10 cm]{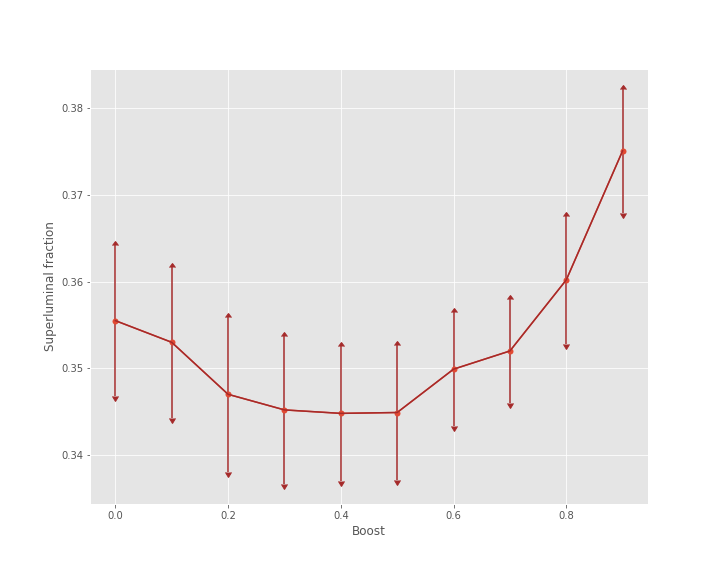}}
 \caption{$P_{super}(v)$ is plotted asa function of the boost velocity for the case of $(\mu_{1},\mu_{2})=\pm 0.5c$ and $\sigma=0.1$ in the lab frame. Average sampling error across all boosts is around 2.2\%.}
 \label{Dwaveplot}
\end{figure}

Now we are in a position to construct a  wavepacket by superposition of plane wave solution of the Dirac equation such that $(\eta \ ,\ \chi)=(\Sigma_{u} c_{u}\exp(i\gamma_{u})\sqrt{1+u} ,\ \Sigma_{u} c_{u}\exp(i\gamma_{u})\sqrt{1-u})$ with an appropriate normalization imposed on it. On substituting this expression in Eq.~[\ref{hj}], we can now obtain a distribution for the weak value for the group velocity operator where the pre-selected states is taken to be a wavepacket with random distribution for $\mu_k$ while the post-selected states are taken from a uniformly random distributed over the Bloch sphere. The  resulting distribution has been plotted in Fig.~[\ref{Dwaveplot}]. The distribution uses the same construction given in Eq.~[\ref{delta}] for the pre-selected state (i.e., averaging over $\mu_k$) but it also incorporates uniform averaging over $\theta$ and $\phi$ in the post-selected state which is what makes it different from the case of K-G waves. The probability density function for Dirac waves is thus given by
 \begin{equation}
     P(v)=\frac{1}{4\pi} \int_{0}^{2\pi} d\varphi \int_{0}^{\pi} d\theta \bra\bra\delta(v-v(x,t))\ket\ket_{\mu_k}.
 \end{equation}
 
In this case the numerical analysis is carried out in a fashion which is very similar to the KG case except for the additional complication coming from the Bloch sphere averaging of the wave-functions. Thus, in order to get the post-averaged distribution of group velocities,  we compute the group velocity at a certain position and time when 200 waves are evolved under the Dirac equation. Relative phases between these components are chosen from a uniform random distribution on $[0,2\pi]$. This calculation of group velocity is then repeated for 5000 values of $\theta$ and $\phi$ along with the 5000 different sets of phases. The distribution formed by these 5000 group velocities form the post-measurement distribution and the superluminal fraction is given by the number of times this weak value is greater than 1. \\

Fig.~[\ref{Dwaveplot}] shows the superluminal probability corresponding to the above situation as a function of the boost velocity. Unlike the case of K-G equation where the fraction is a convex function of the boost, Dirac waves create a concave function. Also the percentage variation of $P_{super}$ is much lower compared to K-G which can be attributed to the two component nature of the wavefunction in the Dirac case (as opposed to the K-G case) and the related uniform averaging scheme over the Bloch sphere.\\

\begin{figure}
 \centerline{\includegraphics[width=10 cm]{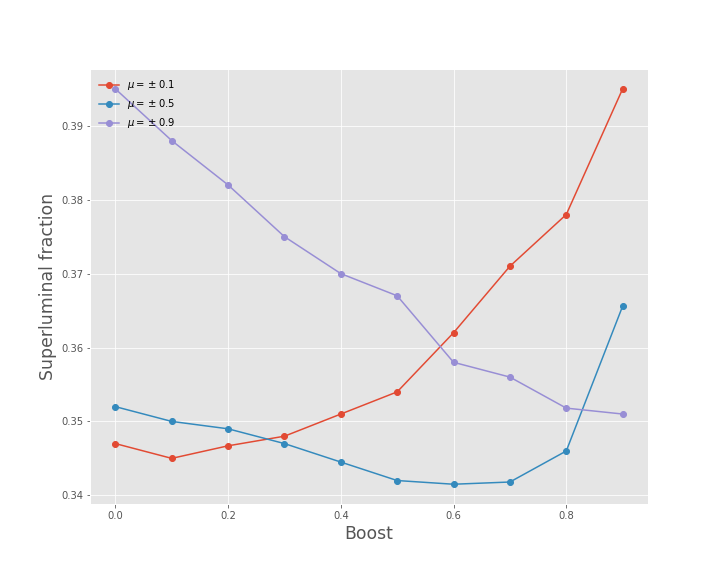}}
 \caption{$P_{super}(v)$ is plotted as a function of the boost velocity for the case of $\mu_{1},\mu_{2}=\pm 0.1c,\pm0.5c,\pm0.9c$ and  $\sigma=0.01$ in the lab frame.. }
 \label{Dsymm}
\end{figure}

\begin{figure}
 \centerline{\includegraphics[width=10 cm]{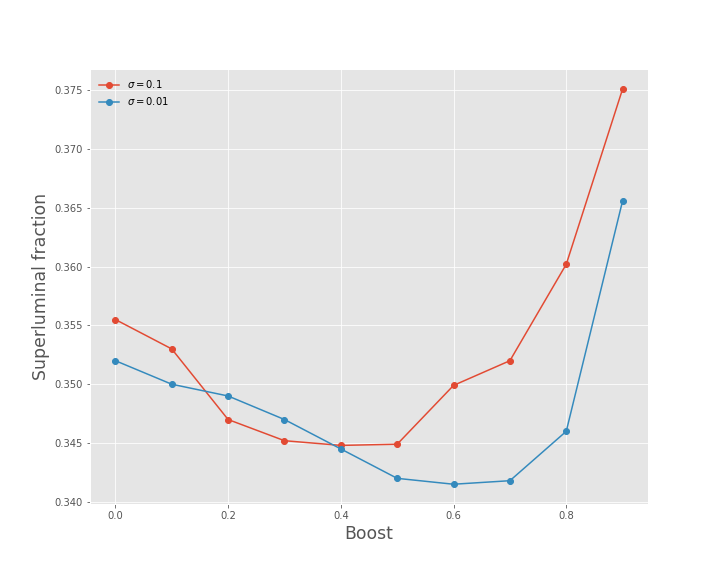}}
 \caption{$P_{super}(v)$ plotted as a function of the boost velocity for the case of $\sigma=0.1$ and $\sigma=0.01$ ($\mu_{1}=0.5c$ and $\mu_{2}=-0.5c$) in the lab frame.}
 \label{Dvar}
\end{figure}

Now we study the influence of various parameters in the spectrum of the pre-selected wave packet prepared in the lab frame which could influence superluminality, i.e., $P_{super}$  in the boosted frame. Unlike the case of K-G, we note that if we consider a pre-selected wavepacket which comprises of more number of partial waves which lay closer to the velocity of light in the lab frame, 
in general, we do not observe enhancement of  superluminality in the boosted frame as can be seen from Fig~\ref{Dsymm}. We also look at the influence of variance of the spectrum in lab frame on $P_{super}$  evaluated in the boosted frame. If we choose a much narrower spectrum such as $\sigma=0.01$, i.e, the waves move with more densely packed wavenumbers, we see that similar to the previous study, the variation of variance in the lab frame has much less of an influence on the change in $P_{super}$ as a  function of boost in comparison to K-G as can be seen from Fig.~\ref{Dvar}.

\section{Discussions and Conclusions:}\label{Discussion} 
In the numerical analysis presented above we noted that the boost can lead to enhancement of superluminality of the weak value in case of both Klein Gordon and Dirac waves. This can be understood as follows. For the K-G case Berry \cite{berry} has shown the distribution of weak values is given by universal form:
\begin{equation}
    P_{super}^{K-G} = {\bigg {[}}\int_{-\infty}^{-1} + \int_{1}^{\infty}{\bigg {]}} ~\frac{\sigma^2}{2((u-\bar{u})^2+\sigma^2)^{3/2}} ~du ,
    \label{bberry}
\end{equation}

\begin{figure}
\centerline{\includegraphics[width=9cm]{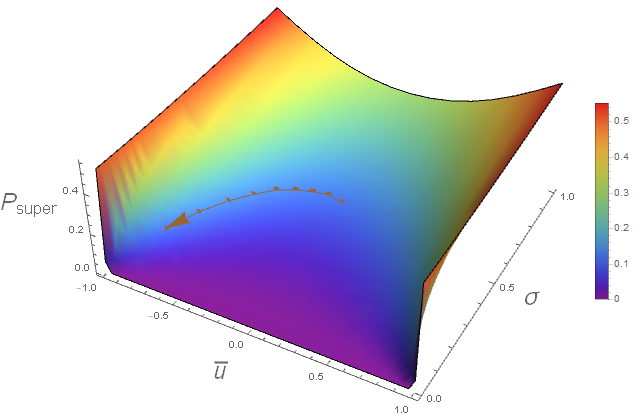}}
 \caption{$P_{super}^{K-G}$ is plotted as a function of $\bar{u}$ and $\sigma$ in the 3D plot. The trajectory of  $P_{super}^{K-G}$ as a function of boost velocity $v$ corresponding to the initial spectrum presented 
 in the Fig.~\ref{KGboost} is superposed on the 3D plot. The arrow shows the direction of evolution of the probability from the lab frame to the boosted frame.}
\label{plot}
\end{figure}
\begin{figure}
\centerline{\includegraphics[width=10cm]{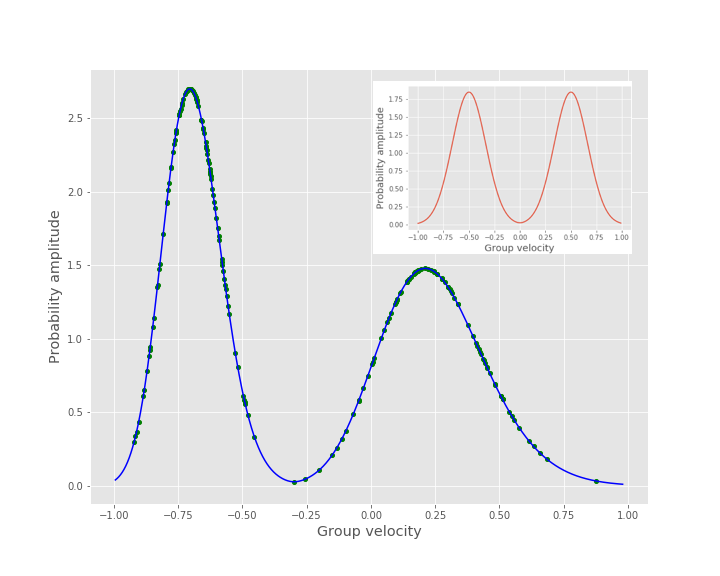}}
 \caption{The group velocity spectrum in the boosted frame using Eq.~[\ref{transformation}] at $v=0.3c$ (in blue). The green dots on the blue curve correspond to the probability amplitude of the wave components in the pre-selected wave function under boost. The corresponding group velocity spectrum  in the rest frame is chosen to be a double gaussian with  $\mu_1,\mu_2 = \pm 0.5c$, $\sigma=0.01$ plotted in red in the inset.}
\label{nonlinear}
\end{figure}
which depends only on the mean ($\bar{u}$)  and standard deviation ($\sigma$) of the  spectrum of `$u$' of the partial waves contributing to the formation of pre-selected wave packet and is independent of the other details. When we apply a boost to the pre-selected wavepacket, it results in transformation of the spectrum of $u$ following Eq.~\ref{transformation}. This in turn results in $\bar{u}$ and  $\sigma$ becoming functions of the boost velocity $v$. And this dependence of $\bar{u} (v)$ and  $\sigma (v)$ on the boost velocity $v$ results in variation of the superluminal probability,  $P_{super}^{K-G}$, as a function of $v$  in accordance with  Eq.~[\ref{bberry}]. The above discussed point is presented graphically in Fig.~[\ref{plot}] where we plot the evolution of $P_{super}^{K-G}$ (for a particular $\bar{u}$ and $\sigma$), under the action of boost as a curve superposed on a surface formed by the universal form of  $P_{super}^{K-G}$ for all values of $\sigma \in (0,1)$ and $\bar{u} \in(-1,1)$. The starting point of the evolution of the curve in Fig.~[\ref{plot}]
corresponds to moments of the initial group velocity distribution in the lab frame as considered in Fig.~[\ref{KGboost}]. The variation (enhancement or suppression) of $P_{super}^{K-G}$ under the action of boost can, therefore be understood completely in terms of a trajectory traced by the evolution of $P_{super}^{K-G}$ on the surface defined by Eq.~[\ref{bberry}].  Therefore the boost gives us a recipe for cruising along on the surface created by Eq.~[\ref{bberry}] and interestingly, that can lead to enhancement of superluminality for both K-G and Dirac waves, although differently as the universal distribution function for the two cases are distinct. In light of the above discussion, we can understand all the numerical analysis presented for Dirac case as well, except for the fact that we would need to replace the universal form of  $P_{super}^{K-G}$ by the universal form of $P_{super}^{Dirac}$. \cite{berry}

It is important to note that the derivation of the universal form for $P_{super}^{K-G}$ is valid for $N \rightarrow \infty$ where $N$ is the number of partial waves forming the pre-selected wave packet. In practice we can consider only a finite number of partial waves to perform our numerical analysis. In the boosted frame, the required value of $N$ for a reasonable reproduction of the universal form for  $P_{super}^{K-G}$ might be different from its value in the lab frame owing to the fact that the  distribution functions of $c_u$'s in boosted frame and lab frame are connected by nonlinear function of the boost velocity $v$ given in  Eq.~[\ref{transformation}]. In Fig.~\ref{nonlinear}, we have shown how the distribution in the lab frame given by a simple double Gaussian distorts itself as a function of boost. This would lead to a different curve for $P_{super}^{K-G}$ compared to its its theoretically predicted universal form under the action of boost for given N. Hence if we need to preform a numerical analysis which faithfully represents the universal form for $P_{super}^{K-G}$ in all frames of reference we need to be judicious in making a choice of N which is large enough.

To conclude, the numerical study presented in this article establish the fact that the local group velocities for relativistic random waves can show a significant enhancement of superluminality  under the action of Lorentz boost. Our findings could motivate new direction for designed of experiments exploring superluminal group velocity\cite{absorption} of light.

\section{Acknowledgements}
It is a pleasure to thank Michael Berry for discussions, encouragement and critical comments on the manuscript. We also thank Pragya Shukla and Oded Zilberberg for discussions on weak measurements and weak values. We thank Nirmalya Ghosh for discussions on fast light medium and superluminal weak values. S.D. would like to acknowledge the ARF grant received from IISER Kolkata and the MATRICS grant (MTR/ 2019/001 043) from Science and Engineering Research Board (SERB) for funding.
 \bibliographystyle{apsrev}

\bibliography{ref}

\begin{thebibliography}{19}
\expandafter\ifx\csname natexlab\endcsname\relax\def\natexlab#1{#1}\fi
\expandafter\ifx\csname bibnamefont\endcsname\relax
  \def\bibnamefont#1{#1}\fi
\expandafter\ifx\csname bibfnamefont\endcsname\relax
  \def\bibfnamefont#1{#1}\fi
\expandafter\ifx\csname citenamefont\endcsname\relax
  \def\citenamefont#1{#1}\fi
\expandafter\ifx\csname url\endcsname\relax
  \def\url#1{\texttt{#1}}\fi
\expandafter\ifx\csname urlprefix\endcsname\relax\def\urlprefix{URL }\fi
\providecommand{\bibinfo}[2]{#2}
\providecommand{\eprint}[2][]{\url{#2}}

\bibitem[{\citenamefont{Einstein}(1905)}]{Einstein}
\bibinfo{author}{\bibfnamefont{A.}~\bibnamefont{Einstein}},
  \bibinfo{journal}{Annalen der Physik} \textbf{\bibinfo{volume}{322}},
  \bibinfo{pages}{891} (\bibinfo{year}{1905}).

\bibitem[{\citenamefont{Brillouin}(1960)}]{Brillouin}
\bibinfo{author}{\bibfnamefont{L.}~\bibnamefont{Brillouin}},
  \emph{\bibinfo{title}{Wave Propagation and Group Velocity}}
  (\bibinfo{publisher}{New York, Academic Press}, \bibinfo{year}{1960}).

\bibitem[{\citenamefont{Sommerfeld}(1914)}]{Sommerfeld}
\bibinfo{author}{\bibfnamefont{A.}~\bibnamefont{Sommerfeld}},
  \bibinfo{journal}{Annalen der Physik} \textbf{\bibinfo{volume}{349}},
  \bibinfo{pages}{177} (\bibinfo{year}{1914}).

\bibitem[{\citenamefont{Born}(1970)}]{Born}
\bibinfo{author}{\bibfnamefont{E.}~\bibnamefont{Born}, \bibfnamefont{M.~Wolf}},
  \emph{\bibinfo{title}{Principles of Optics}} (\bibinfo{publisher}{Pergamon
  Press}, \bibinfo{year}{1970}).

\bibitem[{\citenamefont{Stenner et~al.}(2003)\citenamefont{Stenner, Gauthier,
  and Neifeld}}]{Stenner}
\bibinfo{author}{\bibfnamefont{M.}~\bibnamefont{Stenner}},
  \bibinfo{author}{\bibfnamefont{D.}~\bibnamefont{Gauthier}}, \bibnamefont{and}
  \bibinfo{author}{\bibfnamefont{M.}~\bibnamefont{Neifeld}},
  \bibinfo{journal}{Nature} \textbf{\bibinfo{volume}{425}},
  \bibinfo{pages}{695} (\bibinfo{year}{2003}).

\bibitem[{\citenamefont{Chiao et~al.}(1993)\citenamefont{Chiao, Kwiat, and
  Steinberg}}]{Chiao}
\bibinfo{author}{\bibfnamefont{R.~Y.} \bibnamefont{Chiao}},
  \bibinfo{author}{\bibfnamefont{P.~G.} \bibnamefont{Kwiat}}, \bibnamefont{and}
  \bibinfo{author}{\bibfnamefont{A.~M.} \bibnamefont{Steinberg}},
  \bibinfo{journal}{Sci. Am.} \textbf{\bibinfo{volume}{52(8)}},
  \bibinfo{pages}{269} (\bibinfo{year}{1993}).

\bibitem[{\citenamefont{Steinberg}(1994)}]{Steinberg}
\bibinfo{author}{\bibfnamefont{A.}~\bibnamefont{Steinberg}},
  \bibinfo{journal}{Journal de Physique I, EDP Sciences}
  \textbf{\bibinfo{volume}{4}}, \bibinfo{pages}{1813} (\bibinfo{year}{1994}).

\bibitem[{\citenamefont{Yun-ping and Dian-lin}(1995)}]{Steinberg1}
\bibinfo{author}{\bibfnamefont{W.}~\bibnamefont{Yun-ping}} \bibnamefont{and}
  \bibinfo{author}{\bibfnamefont{Z.}~\bibnamefont{Dian-lin}},
  \bibinfo{journal}{Phys. Rev. A} \textbf{\bibinfo{volume}{52}},
  \bibinfo{pages}{2597} (\bibinfo{year}{1995}).

\bibitem[{\citenamefont{Lee and Lee}(1995)}]{Lee}
\bibinfo{author}{\bibfnamefont{B.}~\bibnamefont{Lee}} \bibnamefont{and}
  \bibinfo{author}{\bibfnamefont{W.}~\bibnamefont{Lee}},
  \bibinfo{journal}{Superlattices and Microstructures}
  \textbf{\bibinfo{volume}{18}}, \bibinfo{pages}{177} (\bibinfo{year}{1995}).

\bibitem[{\citenamefont{Diener}(1996)}]{Diener}
\bibinfo{author}{\bibfnamefont{G.}~\bibnamefont{Diener}},
  \bibinfo{journal}{Physics Letters A} \textbf{\bibinfo{volume}{223}},
  \bibinfo{pages}{327} (\bibinfo{year}{1996}).

\bibitem[{\citenamefont{Lee}(1996)}]{Lee1}
\bibinfo{author}{\bibfnamefont{B.}~\bibnamefont{Lee}}, \bibinfo{journal}{O.S.A.
  Annual Meeting Abstracts (unpublished)} p. \bibinfo{pages}{185}
  (\bibinfo{year}{1996}).

\bibitem[{\citenamefont{Aharonov et~al.}(2003)\citenamefont{Aharonov, Erez, and
  Reznik}}]{aharonovv}
\bibinfo{author}{\bibfnamefont{Y.}~\bibnamefont{Aharonov}},
  \bibinfo{author}{\bibfnamefont{N.}~\bibnamefont{Erez}}, \bibnamefont{and}
  \bibinfo{author}{\bibfnamefont{B.}~\bibnamefont{Reznik}},
  \bibinfo{journal}{Journal of Modern Optics} \textbf{\bibinfo{volume}{50}},
  \bibinfo{pages}{1139} (\bibinfo{year}{2003}).

\bibitem[{\citenamefont{Aharonov et~al.}(1988)\citenamefont{Aharonov, Albert,
  and Vaidman}}]{Aharonov}
\bibinfo{author}{\bibfnamefont{Y.}~\bibnamefont{Aharonov}},
  \bibinfo{author}{\bibfnamefont{D.~Z.} \bibnamefont{Albert}},
  \bibnamefont{and} \bibinfo{author}{\bibfnamefont{L.}~\bibnamefont{Vaidman}},
  \bibinfo{journal}{Phys. Rev. Lett.} \textbf{\bibinfo{volume}{60}},
  \bibinfo{pages}{1351} (\bibinfo{year}{1988}).

\bibitem[{\citenamefont{Berry et~al.}(2011)\citenamefont{Berry, Brunner,
  Popescu, and Shukla}}]{BerryNutrino}
\bibinfo{author}{\bibfnamefont{M.~V.} \bibnamefont{Berry}},
  \bibinfo{author}{\bibfnamefont{N.}~\bibnamefont{Brunner}},
  \bibinfo{author}{\bibfnamefont{S.}~\bibnamefont{Popescu}}, \bibnamefont{and}
  \bibinfo{author}{\bibfnamefont{P.}~\bibnamefont{Shukla}},
  \bibinfo{journal}{J. Phys. A: Math. Theor.} \textbf{\bibinfo{volume}{44}},
  \bibinfo{pages}{492001} (\bibinfo{year}{2011}).

\bibitem[{\citenamefont{Berry}(2012)}]{berry}
\bibinfo{author}{\bibfnamefont{M.~V.} \bibnamefont{Berry}},
  \bibinfo{journal}{Journal of Physics A: Mathematical and Theoretical}
  \textbf{\bibinfo{volume}{45}}, \bibinfo{pages}{185308}
  (\bibinfo{year}{2012}).

\bibitem[{\citenamefont{Berry and Shukla}(2010)}]{berry2}
\bibinfo{author}{\bibfnamefont{M.~V.} \bibnamefont{Berry}} \bibnamefont{and}
  \bibinfo{author}{\bibfnamefont{P.}~\bibnamefont{Shukla}},
  \bibinfo{journal}{Journal of Physics A: Mathematical and Theoretical}
  \textbf{\bibinfo{volume}{43}}, \bibinfo{pages}{354024}
  (\bibinfo{year}{2010}).

\bibitem[{\citenamefont{Clerk et~al.}(2010)\citenamefont{Clerk, Devoret,
  Girvin, Marquardt, and Schoelkopf}}]{Clerk}
\bibinfo{author}{\bibfnamefont{A.~A.} \bibnamefont{Clerk}},
  \bibinfo{author}{\bibfnamefont{M.~H.} \bibnamefont{Devoret}},
  \bibinfo{author}{\bibfnamefont{S.~M.} \bibnamefont{Girvin}},
  \bibinfo{author}{\bibfnamefont{F.}~\bibnamefont{Marquardt}},
  \bibnamefont{and} \bibinfo{author}{\bibfnamefont{R.~J.}
  \bibnamefont{Schoelkopf}}, \bibinfo{journal}{Rev. Mod. Phys.}
  \textbf{\bibinfo{volume}{82}}, \bibinfo{pages}{1155} (\bibinfo{year}{2010}).

\bibitem[{\citenamefont{Dressel et~al.}(2014)\citenamefont{Dressel, Malik,
  Miatto, Jordan, and Boyd}}]{Dressel}
\bibinfo{author}{\bibfnamefont{J.}~\bibnamefont{Dressel}},
  \bibinfo{author}{\bibfnamefont{M.}~\bibnamefont{Malik}},
  \bibinfo{author}{\bibfnamefont{F.~M.} \bibnamefont{Miatto}},
  \bibinfo{author}{\bibfnamefont{A.~N.} \bibnamefont{Jordan}},
  \bibnamefont{and} \bibinfo{author}{\bibfnamefont{R.~W.} \bibnamefont{Boyd}},
  \bibinfo{journal}{Rev. Mod. Phys.} \textbf{\bibinfo{volume}{86}},
  \bibinfo{pages}{307} (\bibinfo{year}{2014}).

\bibitem[{\citenamefont{Brunner et~al.}(2004)\citenamefont{Brunner, Scarani,
  Wegm\"uller, Legr\'e, and Gisin}}]{absorption}
\bibinfo{author}{\bibfnamefont{N.}~\bibnamefont{Brunner}},
  \bibinfo{author}{\bibfnamefont{V.}~\bibnamefont{Scarani}},
  \bibinfo{author}{\bibfnamefont{M.}~\bibnamefont{Wegm\"uller}},
  \bibinfo{author}{\bibfnamefont{M.}~\bibnamefont{Legr\'e}}, \bibnamefont{and}
  \bibinfo{author}{\bibfnamefont{N.}~\bibnamefont{Gisin}},
  \bibinfo{journal}{Phys. Rev. Lett.} \textbf{\bibinfo{volume}{93}},
  \bibinfo{pages}{203902} (\bibinfo{year}{2004}).

\end{thebibliography}

\end{document}